\documentclass[%
reprint,
amsmath,amssymb,
aps,
prb,
]{revtex4-2}
\usepackage{indentfirst}
\usepackage{graphicx}
\usepackage{amssymb}
\usepackage{amsmath}

\begin{document}

\title{Corner Flat Bands Induced by $d$-Density Wave and Partial Corner State Modification Due to Competition with $d$-Wave Superconductivity}

\author{Junming Lao$^{1}$ and Tao Zhou$^{1,2}$}
\email{tzhou@scnu.edu.cn}

\affiliation{$^1$Guangdong Basic Research Center of Excellence for Structure and Fundamental Interactions of Matter, Guangdong Provincial Key Laboratory of Quantum Engineering and Quantum Materials, School of Physics, South China Normal University, Guangzhou 510006, China\\
	$^2$Guangdong-Hong Kong Joint Laboratory of Quantum Matter, Frontier Research Institute for Physics, South China Normal University, Guangzhou 510006, China}

\begin{abstract}
In the realm of condensed matter physics, higher-order topological insulators and superconductors have become a focal point of research due to their unique gapless boundary states at lower-dimensional boundaries such as corners and edges. This paper delves into the effects of a $d$-density wave (DDW) on first-order topological insulators and the competitive dynamics between DDW and $d$-wave superconductivity (DSC) within these systems. We demonstrate that a pure DDW state can instigate a transition to a higher-order topological phase, distinct from DSC behavior, as it is influenced by boundary shape and can generate corner flat bands without additional potentials. Furthermore, we investigate the competitive effects between DSC and DDW, which lead to the transfer of partial corner states. The interplay between these states results in the generation of partial corner states with different symmetries by proximity-induced DSC, while DDW disrupts these and creates new corner states at alternative locations. This study provides valuable insights into the manipulation of topological states through the competition of different order parameters, which could be pivotal for future research and applications in the field.
\end{abstract}


\maketitle

\section{Introduction}
Over the past few decades, topological states have emerged as a major focus of research in condensed matter physics, attracting substantial attention \cite{RevModPhys.83.1057}. Topological states, such as the integer- and fractional-quantum Hall effects \cite{klitzing1980new,tsui1982two}, Chern insulators \cite{haldane1988model}, and topological insulators \cite{kane2005quantum,bernevig2006quantum}, have been extensively studied across a variety of physical systems. Over time, research has led to the concept of more complex and novel topological states. Higher-order topological states have been proposed and have attracted broad interest. In a $d$-dimensional system, a $k$th-order topological insulator or superconductor exhibits $(d-k)$-dimensional gapless edge states \cite{schindler2018higher,benalcazar2017electric,benalcazar2017quantized,song2017d,langbehn2017reflection,cualuguaru2019higher}. This implies that, in addition to materials exhibiting gapless states on surfaces or boundaries, there are also materials with gapless states at lower-dimensional features, such as corner states in a two-dimensional system, thereby significantly expanding the scope of topological state research and its potential applications.

The identification of higher-order topological states has not only enriched the theoretical framework of topological physics but has also been preliminarily proposed in real materials such as 3D bismuth \cite{schindler2018higher_bismuth}, graphene \cite{sheng2019two}, twisted-angle graphene \cite{park2019higher}, and superconductor-topological insulator heterostructures \cite{wang2018high,yan2018majorana,liu2018majorana,li2021rotational,zhu2019second}. Moreover, these states have been explored in various artificial systems \cite{xue2019acoustic,serra2018observation,ni2019observation,peterson2018quantized}. The gapless edge states in these systems are characterized by topological invariants, including mirror Chern numbers \cite{schindler2018higher} and the $Z_4$ index \cite{xu2019higher}. Additionally, apart from intrinsic higher-order topological states, quadrupole/octupole insulators \cite{benalcazar2017quantized,benalcazar2017electric} represent typical extrinsic higher-order topological states, which possess trivial band topology. Research has found that in the presence of a harmonic potential, second-order topology can transform Majorana corner states into Majorana corner flat bands \cite{kheirkhah2020majorana}.

Higher-order topological states may be realized by adding an additional $d$-wave symmetry order parameter to a first-order topological system~\cite{wang2018high,yan2018majorana,liu2018majorana,li2021rotational}. The sign change of the order parameter at the corner of the system results in zero-energy corner states. It has been proposed that higher-order topological superconductors can be realized in heterostructures combining a cuprate superconductor with $d$-wave pairing symmetry and a topological superconductor.

Previously, competing ordered states in unconventional superconductors have garnered considerable attention \cite{doi:10.1126/science.aat4708,zhou2019detecting,lin2021simulating,hu2017topological}. The additional gap induced by competing order parameters can modify topological behavior, facilitating the detection of competing orders via topological properties or the modulation of gap size and Majorana zero modes \cite{xi2015strongly,schmitt2008transient}. The $d$-density wave (DDW) state, a staggered flux state, was originally proposed to account for the pseudogap in cuprate materials~\cite{chakravarty2001hidden}. The DDW order is also proposed to realize higher-order topological states~\cite{wang2022higher}. In the cuprate-based platform, the potential competition between DDW and $d$-wave superconductivity (DSC) could give rise to more complex and intriguing phenomena \cite{li2021rotational}. Beyond real materials, the DDW state may also be realized and controlled in cold atomic systems, providing another platform to realize higher-order topological states \cite{wang2022higher}.

This paper initially investigates the role of DDW in a first-order topological system. The presence of DDW can cause mass variation and band inversion at the corners, transforming gapless boundary states into fully gapped states, thus resulting in the formation of corner states. Unlike DSC-induced corner states, where moderate edge defects do not impact the Majorana corner states and corner flat bands cannot be solely produced by DSC, the corner states formed by DDW are influenced by the strength of DDW and edge defects or shape. Corner flat bands are formed at higher DDW strengths. This paper also explores the potential competition and interplay of DDW order and DSC order in high-T$_c$ superconductor based heterostructures. 

The rest of the paper is organized as follows. In Sec.
II, we introduce the model and present the relevant formalism.
In Sec. III, we report numerical calculations
and discuss the obtained results. Finally, we present the
brief summary in Sec. IV.

\section{Model and Formalism}
We start with the Hamiltonian that includes both the topological insulator part and an additional DDW term, given by:
\begin{equation}
H = H_\mathrm{TI} + H_\mathrm{DDW}.
\end{equation}
Here, \( H_\mathrm{TI} \) describes a two-dimensional topological insulator within the framework of the Bernevig-Hughes-Zhang (BHZ) model \cite{bernevig2006quantum}, expressed as:
\begin{equation}
H_\mathrm{TI}=\sum_{\bf k}C^\dagger_{\bf k} (M_\mathbf{k}\sigma_z +A_x\sin{k_x}\sigma_xs_z+A_y\sin{k_y}\sigma_y + \mu)C_{\bf k},
\end{equation}
where \( C^\dagger_{\mathbf{k}} = (c^\dagger_{a,\mathbf{k}\uparrow}, c^\dagger_{b,\mathbf{k}\uparrow}, c^\dagger_{a,\mathbf{k}\downarrow}, c^\dagger_{b,\mathbf{k}\downarrow}) \). The Pauli matrices \( s_i \) and \( \sigma_i \) act in the spin (\(\uparrow\), \(\downarrow\)) and orbital (\(a\), \(b\)) spaces, respectively. The term \( M_\mathbf{k} = m_0 - t_x \cos k_x - t_y \cos k_y \), and \(\mu\) is the chemical potential.

The Hamiltonian for the DDW term, \( H_\mathrm{DDW} \), is given by:
\begin{equation}
H_\mathrm{DDW}=\sum_{\mathbf{k}\sigma}W_\mathbf{k}(C^\dagger_{\mathbf{k}\sigma}C_{\mathbf{k}+\mathbf{Q}\sigma})
\end{equation}
where \(\mathbf{Q} = (\pi, \pi)\), \( W_\mathbf{k} = 2\mathrm{i} \Delta_\mathrm{DDW} (\cos{k_x} - \cos{k_y}) \), and \(\Delta_\mathrm{DDW}\) denotes the DDW intensity.

The entire Hamiltonian can be recast as an \(8 \times 8\) matrix \(\hat{M}(\mathbf{k})\) with
\begin{equation}
H = \sum_{\mathbf{k}} \mathbf{\Psi}^\dagger(\mathbf{k}) \hat{M}(\mathbf{k}) \mathbf{\Psi}(\mathbf{k}),
\end{equation}
where the vector \(\Psi(\mathbf{k})\) is defined as:
$\Psi_\mathbf{k} = (c_{a,\mathbf{k}\uparrow},$ $c_{b,\mathbf{k}\uparrow}, c_{a,\mathbf{k}\downarrow}, c_{b,\mathbf{k}\downarrow}, c_{a,\mathbf{k}+\mathbf{Q}\uparrow}, c_{b,\mathbf{k}+\mathbf{Q}\uparrow}, c_{a,\mathbf{k}+\mathbf{Q}\downarrow}, c_{b,\mathbf{k}+\mathbf{Q}\downarrow})^T.$

\begin{figure}
	\includegraphics[width=5cm]{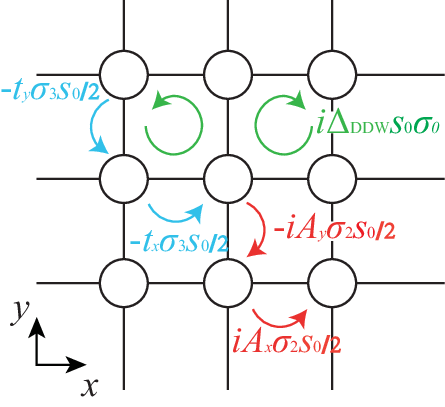}
	\caption{Schematic representation of the lattice model in real space, depicting various parameters in the Hamiltonian.}
	\label{lattice}
\end{figure}

In two-dimensional higher-order topological systems, gapless states are expected to manifest at the corners of the system. To investigate the possible corner states, the Hamiltonian must be expressed in real space through a complete Fourier transformation, as depicted in Fig.~\ref{lattice}.

The Hamiltonian in the real space can be rewritten as,
\begin{equation}
\begin{aligned}
H_\mathrm{TI} = &-\frac{t_\alpha}{2} \sum_{\mathbf{i}, \alpha} \left( C_\mathbf{i}^\dagger \sigma_3  C_{\mathbf{i}+\hat{\alpha}} +{H.c.} \right) + m_0 \sum_\mathbf{i} C_\mathbf{i}^\dagger \sigma_3 C_\mathbf{i} +\mu \\
&- \frac{A_\alpha}{2} \sum_\mathbf{i} \left( i C_\mathbf{i}^\dagger \sigma_1 s_3 C_{\mathbf{i}+\hat{x}} - i C_\mathbf{i}^\dagger \sigma_2 C_{\mathbf{i}+\hat{y}} + {H.c.} \right),
\end{aligned}
\end{equation}
and
\begin{equation}
\begin{aligned}
H_\mathrm{DDW} = &\sum_\mathbf{i} (-1)^{\mathbf{i}_x+\mathbf{i}_y}i\Delta_\mathrm{DDW} C_\mathbf{i}^\dagger C_\mathbf{{i} +\hat{x}}\\
&-\sum_\mathbf{i} (-1)^{\mathbf{i}_x+\mathbf{i}_y}i\Delta_\mathrm{DDW} C_\mathbf{i}^\dagger C_\mathbf{{i}+\hat{y}}.
\end{aligned}
\end{equation}
Here, the site \(\mathbf{j}\) is the nearest neighbor site to the site \(\mathbf{i}\). The vector \( C_\mathbf{i} \) is expressed as \( C_\mathbf{i} = (c_{\mathbf{i}a\uparrow}, c_{\mathbf{i}b\uparrow}, c_{\mathbf{i}a\downarrow}, c_{\mathbf{i}b\downarrow})^T \).

The second-order topological insulator can be characterized by the quadrupole moment \cite{kang2019many,roy2019antiunitary,wheeler2019many,li2020topological}, defined as follows,
\begin{equation}
q_{xy}=\frac{1}{2\pi}\mathrm{Im}\ln{\left[\det(U^\dagger\hat{Q}U)\sqrt{\det(\hat{Q}^\dagger)}\right]},
\end{equation}
where \( Q \equiv \exp{(\mathrm{i} 2\pi \hat{q}_{xy})} \). The operator \(\hat{q}_{xy}\) is the position operator and can be represented as a diagonal matrix with elements being \( \frac{i_x i_y}{N} \). The matrix \( U \) is constructed by column-wise packing of all occupied eigenstates.

The Green's function can be defined by diagonalizing the Hamiltonian matrix \(\hat{M}\). The matrix elements of the Green's function are given by,
\begin{equation}
G(\omega)_{\mathbf{ij}} = \sum_n \frac{c_{\mathbf{i}n} c_{\mathbf{j}n}^\dagger}{\omega - E_n + \mathrm{i} \Gamma}.
\end{equation}

The local density of states (LDOS) is calculated using the real-space Green's function, with
\begin{equation}
\rho_i(\omega) = -\frac{1}{\pi} \sum_{p=1}^4 \mathrm{Im} G(\omega)_{l+p, l+p},
\end{equation}
where \( l = 4(\mathbf{i}-1) \).

The competition between the DSC state and the DDW state is a topic of considerable interest. In systems where DSC is induced by the proximity effect and is influenced by DDW, the phenomenon of corner state migration can occur. Following the approach outlined in Ref.~\cite{li2021rotational}, the Hamiltonian is described as:
\begin{equation}
H = H_\mathrm{TI} + H_\mathrm{SC} + H_\mathrm{DDW} + H_\mathrm{I}.
\end{equation}
The additional terms $H_\mathrm{SC}$ and $H_\mathrm{I}$ represent the superconducting layer and the coupling between the two-dimensional topological insulator and superconductor layers, respectively. The superconducting Hamiltonian $H_\mathrm{SC}$ is given by,
\begin{equation}
    \begin{aligned}
H_\mathrm{SC} = &-\frac{t_\alpha}{2} \sum_{\mathbf{i}\alpha s} \left( c_{\mathbf{i}d s}^{\dagger} c_{\mathbf{i}+\hat{\alpha},d s} + {H.c.} \right) - \mu_\mathrm{d} \sum_{\mathbf{i}s} c_{\mathbf{i}d s}^{\dagger} c_{\mathbf{i}d s}\\ 
&+ \sum_{\langle \mathbf{i}\mathbf{j} \rangle} \left( \Delta_{\mathbf{ij}} c_{\mathbf{i}d\uparrow}^{\dagger} c_{\mathbf{j}d\downarrow}^{\dagger} + {H.c.} \right),
\end{aligned}
\end{equation}
and the coupling Hamiltonian $H_\mathrm{I}$ is:
\begin{equation}
H_\mathrm{I} = -t_{\perp} \sum_{\mathbf{i}, s, \sigma} \left( c_{\mathbf{i}s\sigma}^{\dagger} c_{\mathbf{i}sd} + \text{H.c.} \right),
\end{equation}
where $\sigma \in (a,b)$ represents the orbital index of the topological insulator, and the orbital $d$ belongs to the superconducting layer.

The recast Hamiltonian matrix $\hat{M}$ is an $N \times N$ matrix representing the lattice, with $N = 8N_1 + 4N_2$, where $N_1$ and $N_2$ are the numbers of sites in the topological insulator layer and the superconducting layer, respectively. The vector $\Psi$ is expressed as $\Psi = (C_1,C_2,...C_{N_1},D_1,...,D_{N_2},C_1^\dagger,C_2^\dagger,...C_{N_1}^\dagger,D_1^\dagger,...,D_{N_2}^\dagger)$. Here $C_\textbf{i}=(c_{\textbf{i}a\uparrow},c_{\textbf{i}b\uparrow},c_{\textbf{i}a\downarrow},c_{\textbf{i}b\downarrow})$, and $D_\textbf{i}=(c_{\textbf{i}d\uparrow},c_{\textbf{i}d\downarrow})$.

In our numerical calculations, the parameters are set as follows: $t_x = t_y = 1$, $t_\perp = 0.3$, $A_x = A_y = 1$, $m_0 = 0.5$, $\mu = 0$, $\mu_d = 0.35$, and $\Gamma = 0.005$.

\section{Results and Discussion}

\begin{figure}[!htbp]
    \centering
	\includegraphics[width=\linewidth]{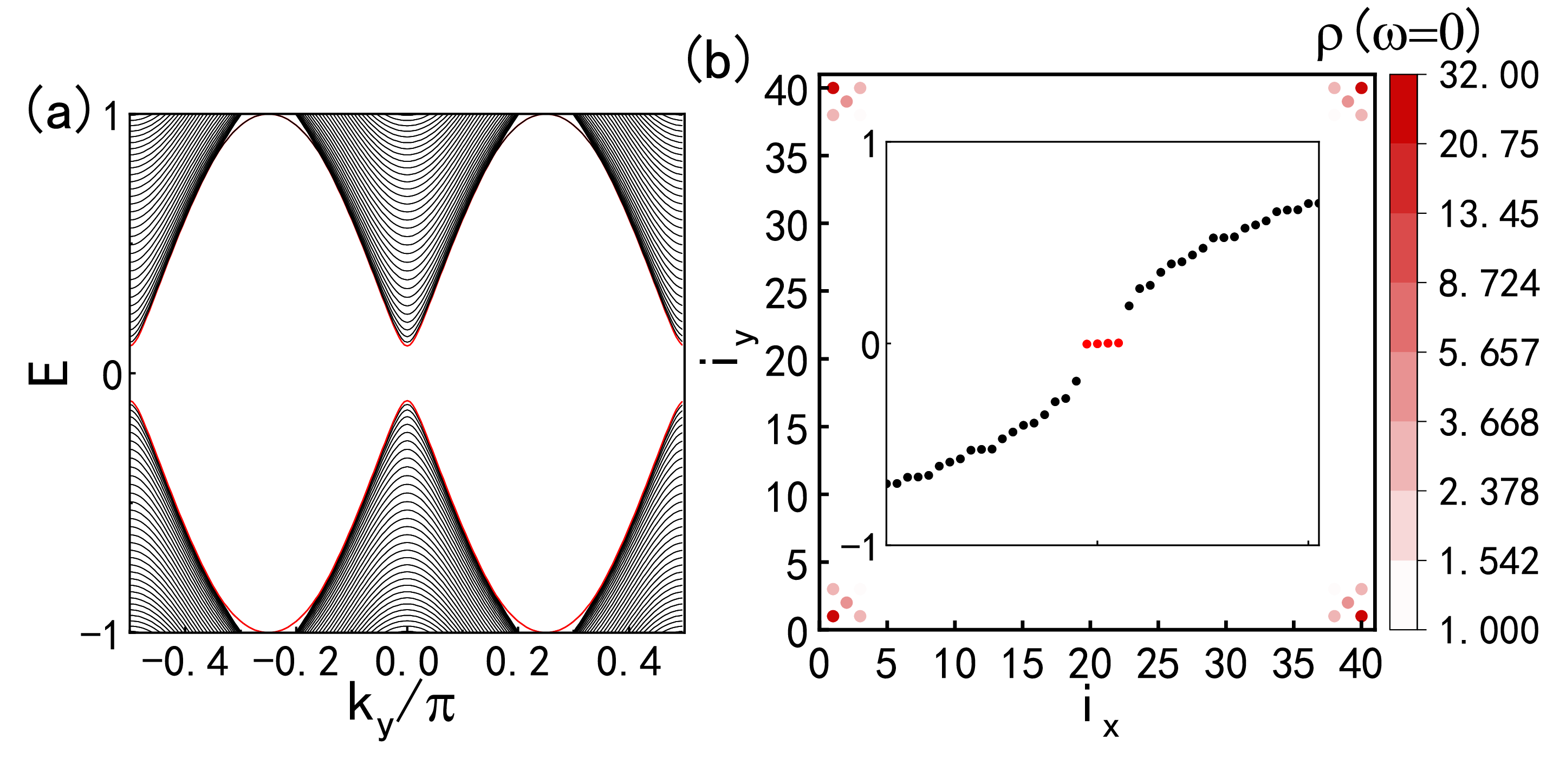}
	\caption{(a) Energy spectrum under open boundary conditions along the $x$-direction for the system with $\Delta_\mathrm{DDW} = 0.15$. The red lines denote the edge states, (b) Intensity of plot zero energy LDOS in the real space with open boundary condition along both $x$ and $y$ directions. Inset of (b): the corresponding eigenvalues of the Hamiltonian, with red dots indicating the topological protected zero energy eigenvalues.}
	\label{xky}
\end{figure}


We first examine the influence of the DDW order on a two-dimensional topological insulator, with its Hamiltonian described by Eq. (1). Under open boundary condition along the $x$-direction and periodic boundary condition along the $y$-direction, the numerical results of the energy bands are shown in Fig.~\ref{xky}(a). As observed, with this cylindrical geometry, the presence of the DDW order results in a fully gapped system with no edge states, indicating that the system does not exhibit first-order topological properties.

Next, we consider open boundary condition along both the $x$- and $y$-directions. The corresponding eigenvalues of the Hamiltonian and the zero-energy LDOS are presented in Fig.~\ref{xky}(b). There are four zero-energy eigenvalues protected by an energy gap. These zero-energy states are located at the corners of the system, suggesting that the system is a second-order topological insulator. This second-order topology can be understood through edge theory~\cite{yan2018majorana}. For a topological insulator described by the BHZ model, the system is expected to be gapless in a cylindrical geometry. However, the addition of a DDW term generally opens an energy gap, with the DDW term acting as an effective mass term. In a square-shaped sample, the sign of the DDW order parameters changes at the corners, leading to the emergence of gapless corner states.

\begin{figure}
	\includegraphics[width=\linewidth]{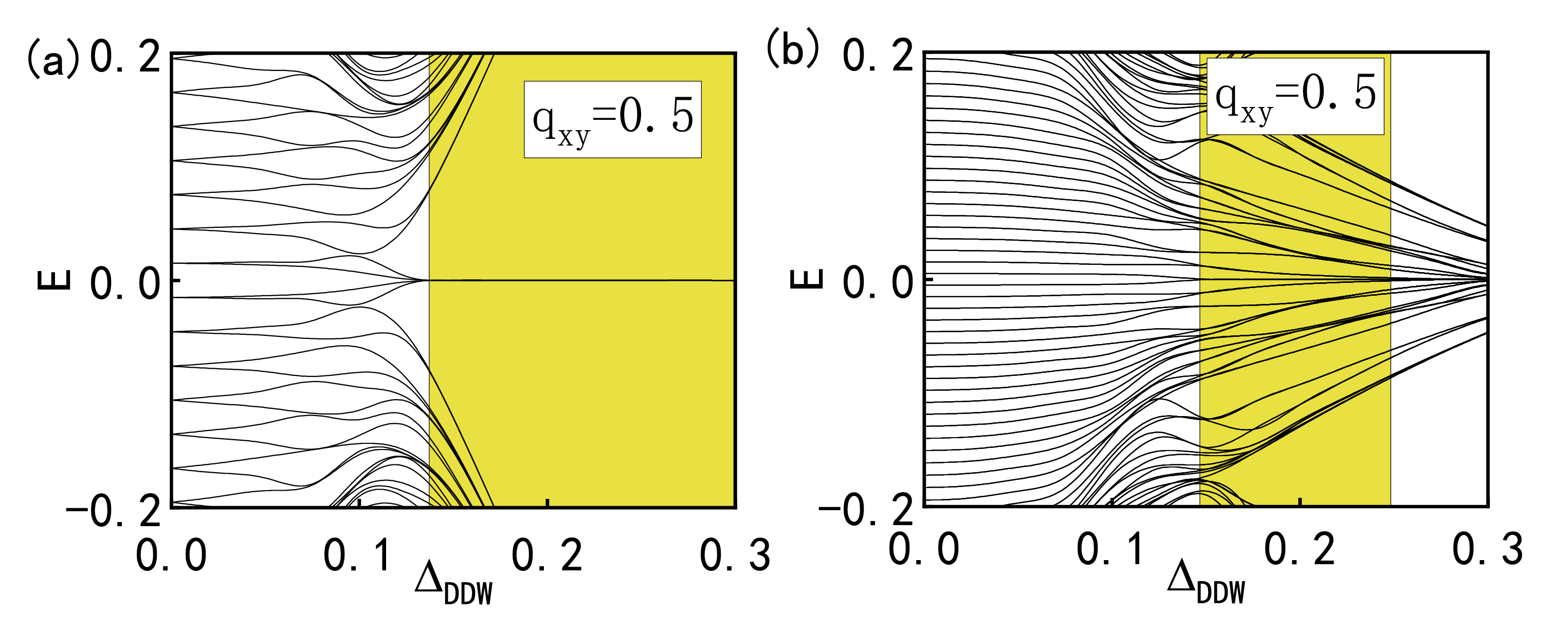}
	\caption{Eigenvalues of the BdG Hamiltonian as a function of $\Delta_\mathrm{DDW}$ for two different lattice geometries: (a) a square lattice and (b) a circular lattice. The yellow region signifies that the system is in a higher-order topological phase, marked by a quadrupole moment of $q_{xy} = 0.5$.}
	\label{square}
\end{figure}

Based on edge theory, the existence of corner states is attributed to the square shape of the sample, where the phase of the order parameter changes at the corners~\cite{yan2018majorana}. It is intriguing to explore the potential edge states under a circular boundary geometry, where the phase of the order parameter changes continuously. Here, the open boundary is considered in both square and circular shapes. The circular shape is defined by $(i_x - 40)^2 + (i_y - 40)^2 < 35^2$, where ${\bf i} = (i_x, i_y)$ represents a lattice site. The eigenvalues as a function of the DDW strength for both square and circular shapes are plotted in Figs.~\ref{square}(a) and ~\ref{square}(b), respectively.

We also calculate the quadrupole moment [Eq.(7)], previously proposed to describe the second-order topology of the system~\cite{kang2019many,roy2019antiunitary,wheeler2019many,li2020topological}. For the square shape, as shown in Fig.~\ref{square}(a), zero-energy eigenvalues emerge when $\Delta_{DDW} \geq 0.14$, protected by an energy gap. Simultaneously, the quadrupole moment $q_{xy}$ remains at $0.5$ when the zero-energy eigenvalues emerge, indicating that the system is indeed a second-order topological insulator and that the quadrupole moment accurately describes the topology.

It has been previously suggested that by introducing a harmonic potential to create an effective circular boundary, the number of corner states in a DSC-induced two-dimensional second-order topological superconductor can be increased. This is achieved by rendering the one-dimensional gapped boundary modes gapless without closing the bulk gap, leading to the emergence of a new type of corner state known as corner flat bands~\cite{kheirkhah2020majorana}. Building on this concept, our research has discovered that the DDW can directly make these one-dimensional gapped boundary modes gapless under circular boundaries, eliminating the need for additional potential.

For the circular shape, as presented in Fig.~\ref{square}(b), the low-energy eigenvalues congregate, and zero-energy eigenvalues emerge as the DDW intensity increases. However, the quadrupole moment $q_{xy}$ equals 0.5 only when $0.15 \leq \Delta_\mathrm{DDW} \leq 0.25$. As $\Delta_\mathrm{DDW}$ increases further, the corner flat bands at zero energy exist, but in this case, the quadrupole moment $q_{xy}$ reduces to zero and cannot describe the topology well. Our results indicate that the second-order topology and the quadrupole moment depend strongly on the shape of the system. The origin of these flat bands is due to the transition of one-dimensional gapped boundary modes into gapless modes, without closing the two-dimensional bulk gap. This transition can be considered a crossover. The wave functions of the corner states and the one-dimensional gapped boundary modes are orthogonal because they are eigenvectors of the Hamiltonian with different energies. As the DDW strength changes, these wave functions gradually evolve, but they remain orthogonal due to the absence of bulk gap closing. Consequently, this crossover stabilizes the corner states by preserving the orthogonality of these wave functions. The closing of the boundary gap does not affect the existence of the corner states, which leads to the formation of the flat band~\cite{kheirkhah2020majorana}.

\begin{figure}
    \includegraphics[width=\linewidth]{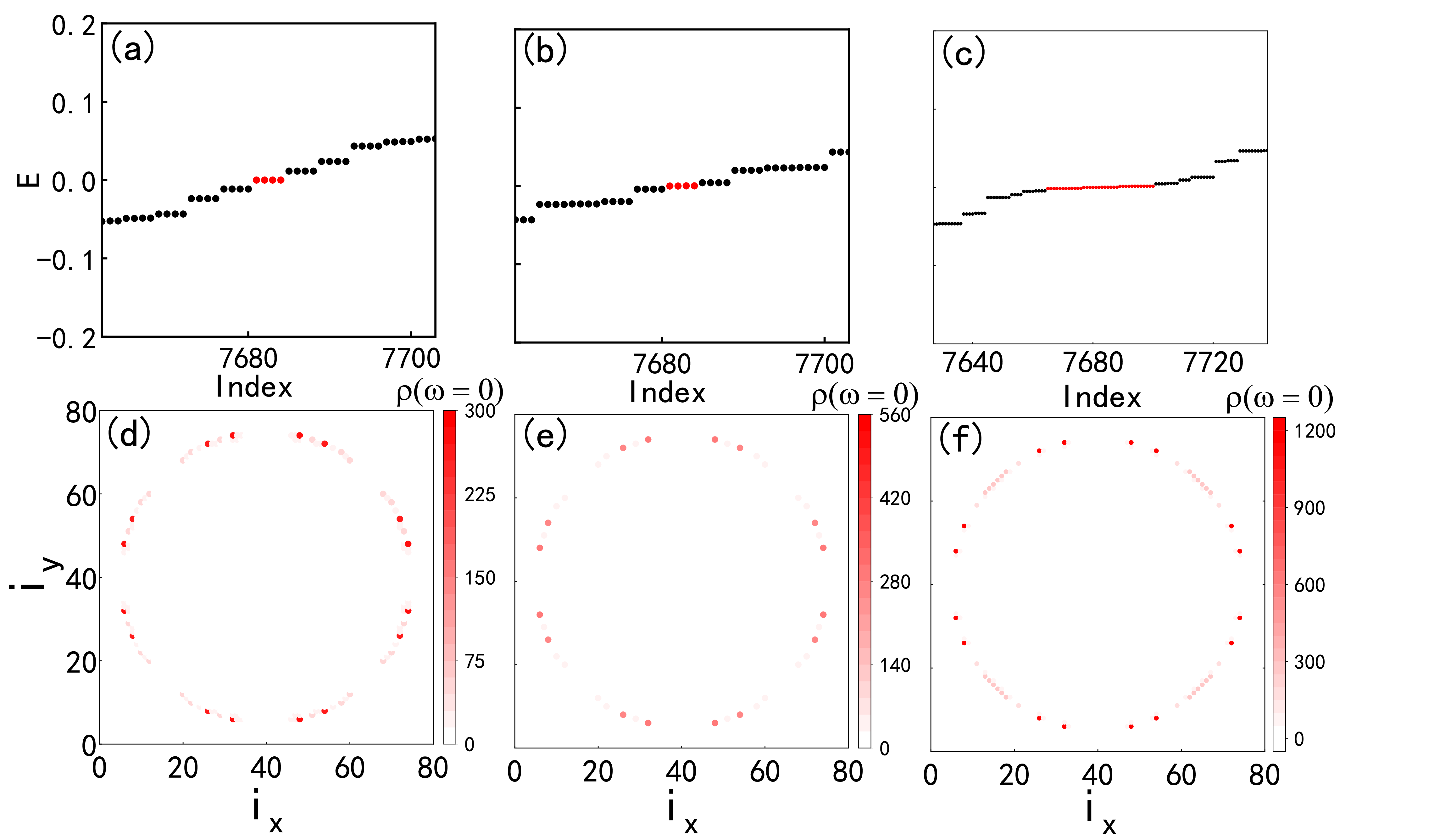}
	\caption{Panels (a)–(c) display the eigenvalues of the BdG Hamiltonian, and panels (d)–(f) present the zero-energy LDOS for a circular lattice with the open boundary condition. The figures correspond to three distinct values of the order parameter: $\Delta_\mathrm{DDW} = 0.15$ for (a) and (d), $\Delta_\mathrm{DDW} = 0.2$ for (b) and (e), and $\Delta_\mathrm{DDW} = 0.3$ for (c) and (f).}
	\label{circular}
\end{figure}

Figs.~\ref{circular}(a)-(c) display the eigenvalues under a circular boundary geometry with $\Delta_\mathrm{DDW} = 0.15, 0.2, 0.3$, respectively. The corresponding zero-energy LDOS are shown in Figs.~\ref{circular}(d)-(f), respectively. Due to the presence of quasiparticle damping, the zero-energy LDOS also includes the excitations of low-energy quasiparticles. As the DDW intensity is small, as is seen in Fig.~\ref{circular}(a), the zero energy states are protected by a small energy gap. In this case, the zero-energy LDOS is completely contributed by the zero energy eigenvalues. As the DDW intersity increases, the low energy eigenvalues congregate to zero energy. The zero flat flat band emerges, as is seen in  
Figs.~\ref{circular}(b) and ~\ref{circular}(c). The zero energy eigenvalues 
continuous connects to some low energy eigenvalues. In this case, the zero energy LDOS presented in Figs.~\ref{circular}(f)
 is not only contributed by zero-energy states but also includes some low-energy contributions.

The origin of the flat bands lies in the closing of low energy edge states rather than bulk states. As shown in Fig.~\ref{square}(b) and Fig.~\ref{circular}(a), when $\Delta_\mathrm{DDW} = 0.15$, the system undergoes a transition from a first-order topological insulator to a second-order topological insulator. The gapless topologically protected states, which were originally in (2-1)-dimensions, transform into topologically protected corner states. When $\Delta_\mathrm{DDW} = 0.2$, the system is in a second-order topological state. Fig.~\ref{circular}(b) shows that the system has distinct corner states, while the one-dimensional boundaries are closing.

Finally, when $\Delta_\mathrm{DDW} = 0.3$, Fig.~\ref{circular}(c) shows that the energy gap of the edge is closed by the DDW. At this point, Fig.~\ref{circular}(f) reveals that other positions along the boundaries also exhibit zero-energy LDOS, indicating that the previously gapped one-dimensional edge modes eventually become gapless, forming a zero energy corner flat band. We have numerically verified that the probability density of the eigenvalues on the flat band is indeed localized at the boundary, and only along the boundary the energy gap is closed.

Moreover, due to the staggered factor of the DDW term coefficient $(-1)^{i_x + i_y}$, which has two patterns set at the one-dimensional boundaries, there are two patterns of zero-energy LDOS that change with DDW. One pattern, as illustrated in Figs.~\ref{circular}(d)-(f), is originally distributed at the left, right, top, and bottom, expanding to the top-left, top-right, bottom-left, and bottom-right as the DDW intensity increases. Conversely, the other pattern starts from the four corners and extends to the left, right, top, and bottom as DDW increases. Since there are no significant differences between the two patterns, aside from these initial distributions, we have not displayed the latter, and our conclusions hold for both cases.

Notably, in the case of DSC-induced second-order topological states, no corner flat bands exist for the circular geometry. For superconducting systems with particle-hole symmetry, unless protected by specific symmetries, multiple Majorana zero modes at the same space will generally annihilate, rendering the system gapped. As a result, for the DSC system, realizing a Majorana corner flat band is challenging. However, as we have mentioned, it has been proposed that this can be achieved by considering an additional harmonic potential~\cite{kheirkhah2020majorana}.

\begin{figure}
	\includegraphics[width=\linewidth]{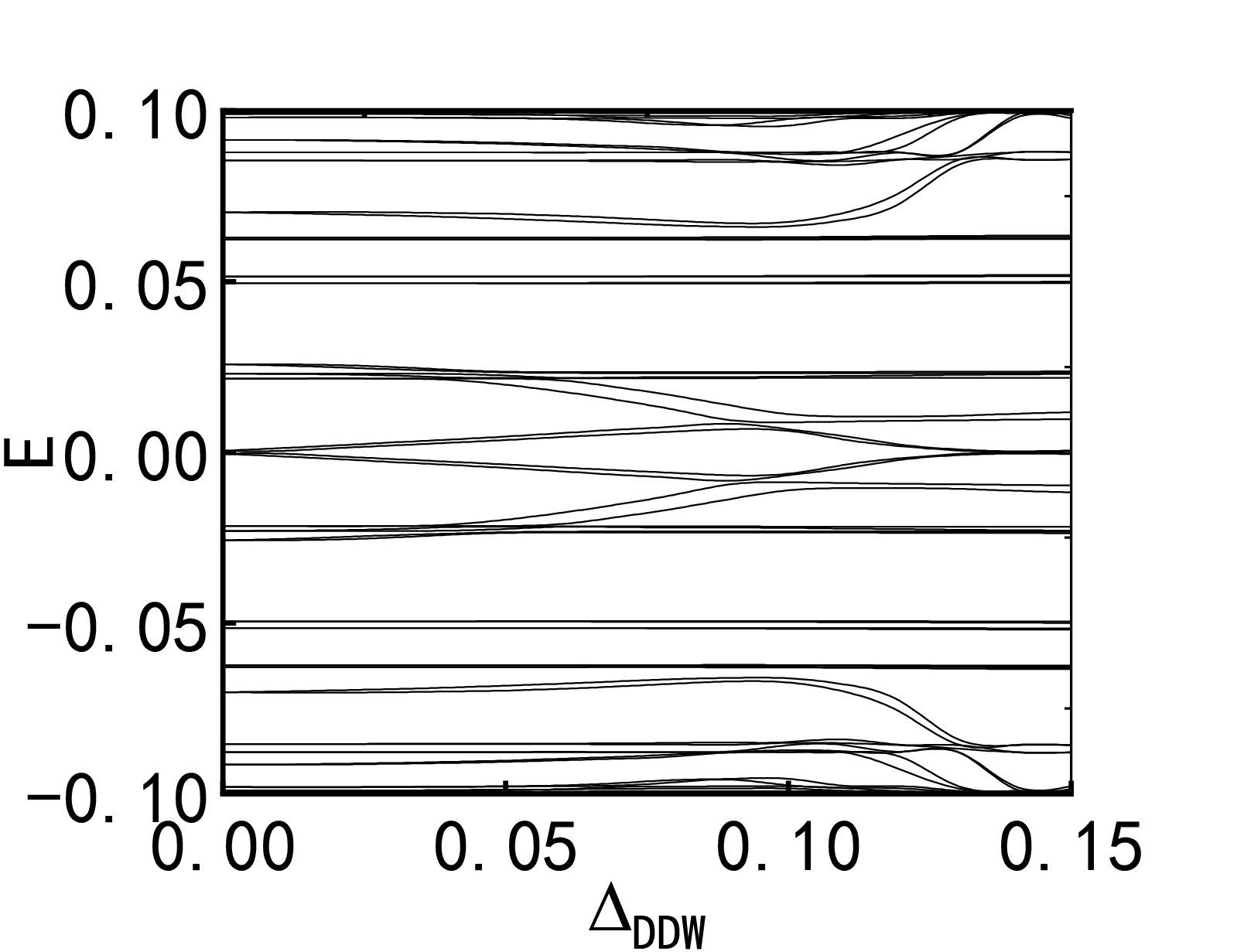}
	\caption{Eigenvalues of the BdG Hamiltonian plotted as a function of $\Delta_\mathrm{DDW}$, with $\Delta_\mathrm{DSC}$ fixed at $0.3$.}
	\label{proximity_ddw_dsc_en}
\end{figure}


Previously, the second-order topological superconductor induced by the DSC order has been studied intensively~\cite{wang2018high,yan2018majorana,liu2018majorana,li2021rotational}. Specifically, it was proposed that zero-energy states only emerge at certain corners within a microscopic model~\cite{li2021rotational}. We have demonstrated that the DDW order can also induce zero-energy corner states, as presented in Fig.~\ref{xky}. It is of interest to explore the influence of the DDW order on the DSC-induced second-order topological superconductor
and investigate the interplay of two different zero-energy corner states.

The combined effect of the DDW order and DSC order on the zero-energy states is illustrated in Fig.~\ref{proximity_ddw_dsc_en}, where the eigenvalues of the Hamiltonian with the square boundary with $N_1=40\times40$ and $ N_2=60\times60$ are presented as a function of the DDW intensity. Without the DDW order, zero-energy states exist, indicating topological behavior. In the presence of the additional DDW order, the zero-energy states disappear. As the DDW intensity increases beyond 0.13, the zero-energy states reappear.

\begin{figure}
	\includegraphics[width=\linewidth]{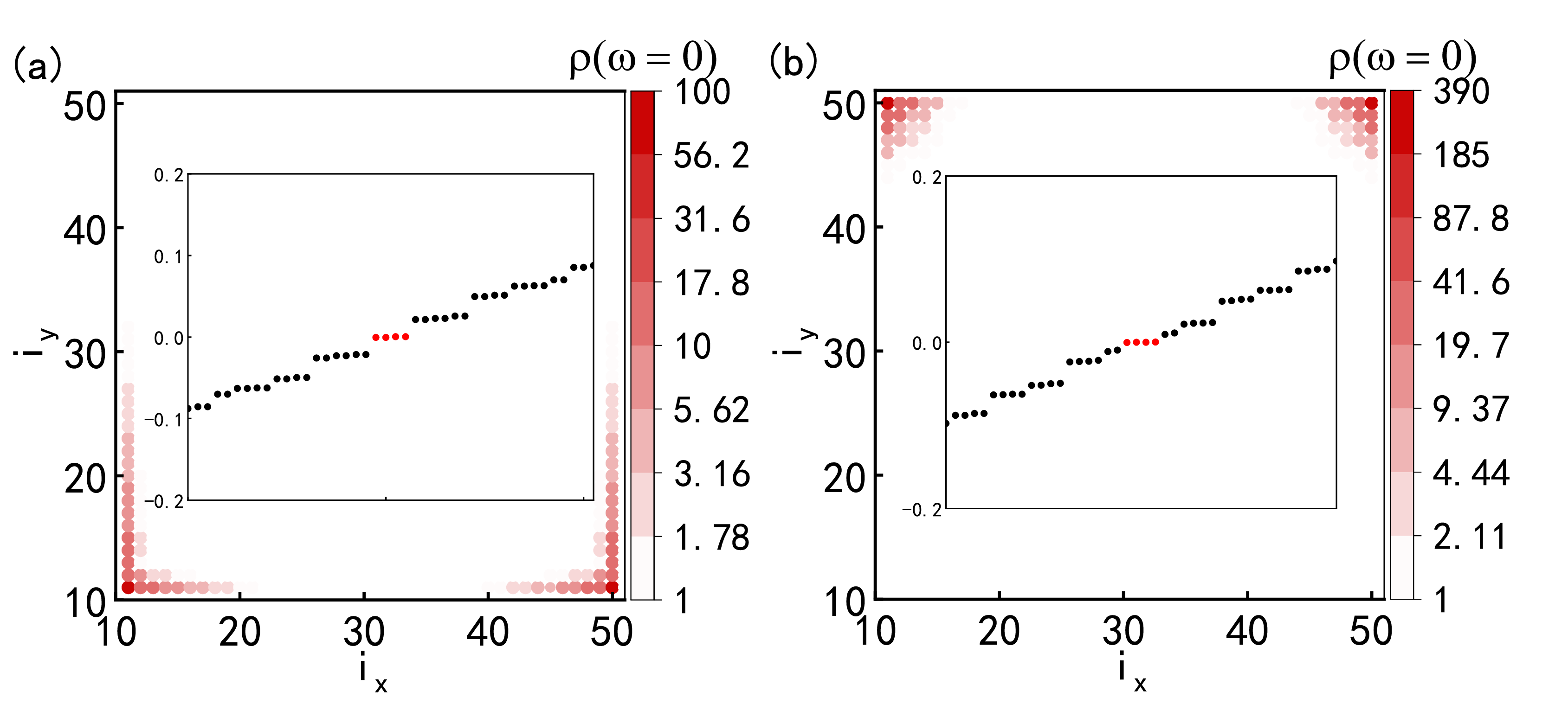}
	\caption{The eigenvalues and zero-energy LDOS for (a) $\Delta_\mathrm{DDW}=0$ and (b) $\Delta_\mathrm{DDW}=0.15$, both with $\Delta_\mathrm{DSC}=0.3$. Here $\Gamma=0.0001$}
	\label{proximity}
\end{figure}

We present the numerical results of the eigenvalues and the zero-energy LDOS without and with the DDW order in Figs.~\ref{proximity}(a) and \ref{proximity}(b), respectively. In the pure DSC state, as shown in Fig.~\ref{proximity}(a), zero-energy states emerge at two bottom corners, originating from the mixing of the band structure and the proximity-induced asymmetric pairing order parameter in the two-dimensional insulator layer. The absence of zero-energy states at the top corners can be explained by exploring the anomalous Green's function at the system edges~\cite{li2021rotational}. In the presence of the DDW order, corner states emerge at the top corners due to the phase change of the DDW order. However, at the bottom corners, the zero-energy states induced by the DSC order and the DDW order merge and annihilate. Therefore, in the presence of the DDW order, the Majorana zero modes shift from the bottom corners to the top corners. 
We here have provided an effective platform to explore the competition for the DSC order and DDW order. Also, in this platform, the DDW order can be used to control and manipulate the zero modes, which may provide potential applications.

Our research yields valuable theoretical insights that are poised to influence future experimental studies. Previous studies have indicated that DSC order and other competing orders might coexist within high-T$_c$ superconductors \cite{doi:10.1126/science.aat4708,zhou2019detecting}. Concurrently, recent breakthroughs have enabled the creation of two-dimensional topological insulators capable of functioning at temperatures up to 100 K \cite{wu2018observation}. Moreover, there has been a successful development of heterostructures that combine high-T$_c$ superconductors with topological insulators \cite{zareapour2012proximity,wang2013fully,wan2020twofold,zhao2018superconducting,chen2018superconductivity}. Consequently, we anticipate that our theoretical proposals could be translated into experimental realities.

Within the realm of high-T$_c$ superconductors, the potential for coexistence and competition among various orders is of paramount importance. The interplay between these competing orders and superconductivity can significantly alter topological properties or be concealed by the superconducting gap \cite{mueller2017review,nayak2000density,chakravarty2001hidden,li2006checkerboard,fradkin2015colloquium}. Unraveling the intricacies of these competitive interactions is crucial not only for advancing our theoretical understanding of unconventional superconductivity but also for directing future experimental research in topological superconductivity. Therefore, it is particularly enlightening to scrutinize the competitive dynamics between the superconducting pairing order and the DDW order. We are optimistic that our current research will lay the groundwork for future scientific breakthroughs.

Expanding beyond traditional materials, the manifestation of DDW states in cold atom systems is also a viable prospect. Quantum simulations of the BHZ Hamiltonian using cold atomic gases have achieved the measurement of spin Berry curvature and spin Chern numbers \cite{lv2021measurement}. These groundbreaking experiments utilized atomic quantum gases to emulate the four-band BHZ model with two pseudospins. Such advancements highlight the feasibility of realizing two-dimensional topological insulators with DDW states in cold atom platforms, offering a promising avenue for further exploration.

\section{Summary}

In summary, this study provides a comprehensive investigation into the effects of DDW on topological insulators. We discovered that DDW can induce the formation of corner flat bands in higher-order topological systems with circular boundaries, a phenomenon not observed in DSC systems, and this can be achieved without the need for additional conditions such as harmonic potentials. Our findings reveal that as the DDW strength increases, the system undergoes a transition from boundary states to corner states, and eventually to corner flat bands.

In the context of higher-order topological superconducting systems, we found that the effective pairing terms induced by tunneling can lead to the emergence of partial corner states with symmetries distinct from the original superconductor. The introduction of a competing DDW order can disrupt these original corner states. However, at higher DDW strengths, new corner states form at various locations. Additionally, we observed that considering the strengths of both DSC and DDW can lead to low-energy modes with varying intensities of corner states.

Our research sheds new light on the role of DDW in topological systems and its potential to manipulate corner states. The findings presented here are expected to inform and guide future research in the field of topological insulators and superconductors.

\begin{acknowledgments}
	This work was supported by the NSFC (Grant No.12074130).
\end{acknowledgments}

\bibliography{main}

\end{document}